\documentclass[preprint]{aastex}

\shorttitle{New high proper motion stars at high galactic latitudes}
\shortauthors{Lepine et al.}

\begin{document}

\title{New High Proper Motion Stars from the Digitized Sky Survey.
II. Northern Stars with $0.5<\mu<2.0\arcsec yr^{-1}$ at High Galactic
Latitudes.\altaffilmark{1}}

\author{S\'ebastien L\'epine\altaffilmark{2,3}, Michael
  M. Shara\altaffilmark{2}, and R. Michael Rich\altaffilmark{4}}

\altaffiltext{1}{Based on data mining of the Digitized Sky Survey,
developed and operated by the Catalogs and Surveys Branch of the Space
Telescope Science Institute, Baltimore, USA.}

\altaffiltext{2}{Department of Astrophysics, Division of Physical Sciences,
American Museum of Natural History, Central Park West at 79th Street,
New York, NY 10024, USA}

\altaffiltext{3}{Kalbfleich Research Fellow}

\altaffiltext{4}{Department of Physics and Astronomy, University of
California at Los Angeles, Los Angeles, CA 90095, USA}

\begin{abstract}
In a continuation of our systematic search for high proper motion
stars in the Digitized Sky Survey, we have completed the analysis
of northern sky fields at galactic latitudes above 25 degrees. 
With the help of our SUPERBLINK software, a powerful automated blink
comparator developed by us, we have identified 1146 stars in the
magnitude range $8<r<20$ with proper motions $0.500<\mu<2.000\arcsec$
yr$^{-1}$. These include 1080 stars previously listed in Luyten's
proper motion catalogs (LHS, NLTT), 9 stars not previously listed in
the Luyten catalogs but reported elsewhere in the literature
(including 1 previously reported by our team), and 57 new objects
reported here for the first time. This paper includes a list of
positions, proper motions, magnitudes, and finder charts for all the
new high proper motion stars. Combined with our previous study of low
galactic latitude fields (see Paper I), our survey now covers over
98\% of the northern sky. We conclude that the Luyten catalogs were
$\simeq$90\% complete in the northern sky for stars with
$0.5<\mu<2.0\arcsec$ down to magnitude r=19. We discuss the
incompleteness of the old Luyten proper motion survey, and estimate
completeness limits for our new survey.
\end{abstract}



\section{Introduction}

One of the main interests of large proper motion surveys is their use
in identifying the low-luminosity stellar components of our
Galaxy. Surveys of faint stars with large proper motions have
traditionally been the most powerful means to identify intrinsically
faint objects in the neighborhood of the Sun. In particular, the
vast majority of all the known nearby red dwarfs, red subdwarfs, and
white dwarfs, have first been identified as high proper motion stars.

The most extensive catalogs of high proper motion stars are the two
catalogs by W. H. Luyten, published over 20 years ago. The {\it
LHS catalogue} (Luyten 1979) lists 3602 objects with estimated
proper motions $\mu>0.500\arcsec$ yr$^{-1}$ and 867 other stars
with estimated proper motions $0.235<\mu<0.500\arcsec$
yr$^{-1}$. The NLTT catalogue (Luyten 1979) is essentially an
extension of the LHS catalogs, and lists 58845 high proper motion
stars, the majority of which have estimated proper motions
$\mu\geq0.18\arcsec$ yr$^{-1}$. The vast majority of the $\approx3000$
solar neighborhood stars are listed as high proper motion stars in the
Luyten catalogs. Furthermore, new nearby stars continue to be
identified among LHS and NLTT stars, the majority of which still do
not have any formal spectroscopic classification.

The LHS and NLTT catalogs are significantly incomplete in the Southern
sky and at low galactic latitudes, leaving large expanses of the Solar
neighborhood that still are very poorly surveyed for intrinsically
faint stars. To remedy this, several new surveys of high proper motion
stars are now being conducted on different parts of the sky. Most of
the new, deep proper motion surveys are being conducted in the
Southern sky. These include the \citet{SIIJM00} survey, the
Cal\'ar-ESO proper motion survey \citep{RWRG01}, and the
Liverpool-Edinburgh proper motion survey \citep{PJH03}. These surveys
are all based on analysis of photographic plate material obtained from
the 1970s to the 1990s (the United Kingdom Schmidt Telescope survey
and the ESO-schmidt survey). Thus far, little attention has been directed
toward the northern sky since the publication of the Luyten catalogs
\citep{MFLCHR00}. However, the northern Palomar Observatory Sky
Surveys (POSS) represent one of the most interesting opportunities for
proper motion investigations. The first and second epoch surveys (the
1950s POSS-I, and the 1990s POSS-II) together provide a large baseline
of 35-45 years which provides better sensitivity to high proper motion
stars, and greater accuracy on proper motion measurements.

The USNO-B1.0 all-sky catalog, recently made available
\citep{Metal03}, constitutes a major achievement in the
characterization of the largest possible number of objects detected in
the large photographic surveys. The northern part of the catalog is
based on scans of the POSS-I and POSS-II plates. Besides providing
relatively good astrometry and optical photometry for an unprecedented
number of stars, the USNO-B1.0 also tentatively lists proper motions,
when detected. However, a recent analysis of the proper motions in the
USNO-B1.0 by \citet{G03} shows that the USNO-B1.0 in its current form
cannot be used to assemble a complete catalog of high proper motion
stars.  The proper motion errors in USNO-B1.0 are relatively large and
many stars known to have large proper motions are not detected as
such. For example, the USNO-B1.0 does not correctly identify as high
proper motion objects more than 10\% of the Luyten stars. Furthermore,
the USNO-B1.0 is too liberal in it identification of moving objects,
and contains some 200$\times$ too many high proper motion star
entries, the vast majority of which are bogus detections.

We have been conducting an independent analysis of the Palomar Sky
Surveys, using a more direct approach to the detection of high proper
motion stars. The SUPERBLINK software, developed by one of us (SL),
works like an automated blink comparator, and uses a difference method
to identify variable and moving objects in pairs of sky images
obtained at two different epochs \citep{LSR02b}(\-- hereafter paper
I). Our method has proved particularly successful in finding moving
objects in densely populated regions, areas that have traditionally
been avoided in proper motion searches. Furthermore, every object
identified by our code is assigned a probability of being real, and
the best candidates are all confirmed visually by direct inspection of
the DSS images. As a result, the number of bogus detections is
extremely low.

In Paper I we have presented the results of a search at low galactic
latitudes ($|b|<25.0$) for stars with proper motions
$0.5<\mu<2.0\arcsec$. Within a search area covering $\approx$9000
square degrees, we discovered 141 previously unreported high proper
motion stars. In this paper, we extend our analysis to northern sky
fields at higher galactic latitudes ($|b|>25.0$). In \S 2, we briefly
describe our identification methods and our assignment of
magnitudes. \S 2 compares our survey with the Luyten proper motion
catalogs. The new stars discovered in our surveys are presented in
detail in \S 4. In \S 5 we use reduced proper motion diagrams to
separate the high proper motion stars into three classes of objects. A
summary and conclusions follow in \S 7.


\section{Proper Motion Search and Identification}

The method used for the identification of high proper motion stars
from Digitized Sky Survey images is described in detail in Paper I. We
provide here only a brief summary.

The survey is based on Digitized Sky Survey (DSS) scans of the first
and second epoch Palomar Observatory Sky Survey (POSS-I, POSS-II) in
the red. The area covered by the POSS-I plates is divided into a grid
of 642,900 square subfields $17\arcmin\times17\arcmin$ in size, with a
$15\arcmin$ grid spacing allowing for a significant overlap. For each
subfield, a pair of DSS images (POSS-I, POSS-II) is retrieved and
analyzed with the SUPERBLINK software. The code performs a
superposition of the two images by minimizing the residuals after
shifting and rotating the POSS-II image onto the POSS-I grid. The
POSS-II red image is also degraded to match the resolution and quality
of the POSS-I red image. The SUPERBLINK software then uses search
algorithm to identify and characterize the differences in the two
images. Variable stars, plates defects, and moving objects are
identified and cataloged. Finders charts of candidate high proper
motion stars, in the form of $4.25\arcmin\times4.25\arcmin$ dual-epoch
images, are generate by the code. These can be blinked on the
computer, and prospective objects are all confirmed by visual
inspection.

Once the existences of high proper motion stars are confirmed visually,
we search for counterparts of the objects in large astrometric and
photometric catalogs. For the current sample of high galactic
latitude, high proper motion stars, we used the USNO-B1.0 catalog
(VizieR catalog I/284) to obtain optical, photographic magnitudes, and
the 2MASS All-Sky Point Source Catalog (PSC) to get infrared
magnitudes. Since the USNO-B1.0 catalog is based on the same plate
material, we do find counterparts for the majority of our confirmed
high proper motion stars, although a small number of confirmed objects
are notlisted in the USNO-B1.0, for reasons that are unclear at this
point. Likewise, we do also find 2MASS counterparts for the majority
of our stars, but a small number of faint objects do not show up,
most likely because they are not detected in the infrared.

To this date, a total of 623,474 subfields, or $\simeq97\%$ of the sky
north of $\delta=-3.0^{\circ}$, have been successfully processed with
SUPERBLINK. Some 4,768 subfields ($\simeq0.74\%$) were rejected by the
code for a variety of reasons, including the presence in the subfield
of a very bright (R$<5$) star, a large plate defect, or an extended,
saturated object such as the Andromeda Galaxy. The remaining 14,658
fields from our grid that have not been processed yet are mainly
fields at very high declinations ($\delta>+88.0$), which we avoided
because of potential problems in superposition, and subfields near the
low declination limit of the POSS-I plates ($-3.0<\delta<-1.0$) which
are not strictly part of the northern sky.

In the area north of $\delta$=0.0$^{\circ}$ analyzed by SUPERBLINK, we
have identified a total of 1797 stars with proper motions
$0.500<\mu<2.000\arcsec$ yr$^{-1}$; 1146 at high galactic latitudes
($|b|>25$), and 651 at low galactic latitudes ($|b|<25$). We have
obtained complete $bri$ magnitudes from the USNO-B1.0 for 1691 of the
stars. Partial information, i.e. a magnitude in at least one of the
photographic bands, was obtained for 64 more stars. We were unable to
find a counterpart in the USNO-B1.0 catalog for the 42 remaining
objects, although they all clearly show up in the DSS. Whenever a
star didn't have a counterpart in the USNO-B1.0 catalog in the $r$ band,
we used the $r$ magnitude estimated by SUPERBLINK. Infrared $JHK_s$
magnitudes were found in the 2MASS All-Sky PSC for 1754 of our high
proper motion stars; the remaining 43 stars most probably are too
faint in the infrared to have been detected by 2MASS.

To estimate the astrometric errors in the positions derived by
SUPERBLINK, we compare with the quoted positions of the 2MASS
counterparts. On output, SUPERBLINK gives the position of the star
at the epoch 2000.0. Given the proper motion vector, we extrapolate
the position of each star at the epoch of the 2MASS observations. The
distribution of errors between the SUPERBLINK and 2MASS positions are
shown in Figure 1. The positions of the 2MASS counterparts, as given by
the 2MASS All-Sky PSC, are accurate to about $0.15\arcsec$
in both RA and DEC. Our distribution of the differences between our
positions and the 2MASS positions suggests there is a mean error of
$\approx1.2\arcsec$ on the absolute positions estimated with
SUPERBLINK. These are admittedly much less accurate than the positions
quoted in the USNO-B1.0 catalog, which are supposed to be accurate to
$0.2\arcsec$. However, our absolute positions are entirely dependent
on the accuracy of the POSS-I plate solutions calculated for the DSS,
which we are using. Because we are shifting and degrading the POSS-II
DSS image to superpose it on the POSS-I DSS image, we effectively lose
much of the information provided by the superior POSS-II plates to
determine absolute positions. In addition, most of the stars for which the
SUPERBLINK absolute positions have errors $>2\arcsec$ are relatively
bright objects ($r<14$) that have a saturated core on the POSS-I and
POSS-II plates, and whose absolute position cannot be determined very
accurately using the SUPERBLINK algorithms.

Actually, SUPERBLINK has been specifically designed to measure 
{\it relative proper motions} between two images, and not the absolute
position of the stars. The relative position of a moving star between
the POSS-I and POSS-II images is generally calculated to better than
0.4 arcsec, which yields an accuracy on the proper motion better than
$0.01\arcsec$ yr$^{-1}$. This can be verified by comparing the proper
motions calculated for the LHS stars by SUPERBLINK with the proper
motions given in the USNO-B1.0 catalog. The USNO-B1.0 catalog lists
reasonably accurate proper motions for $\approx85\%$ of the 1199 high
proper motion stars identified with SUPERBLINK at high galactic
latitudes. The other $15\%$ of the stars have proper motions quoted in
the USNO-B1.0 that are clearly inaccurate, and most of them are
actually listed as not having any measurable proper motion. Figure 2
shows the distribution of the difference between the proper motions
estimated with SUPERBLINK, and those given in the USNO-B1.0 catalog.
The difference is generally smaller than $\pm0.01\arcsec$ yr$^{-1}$,
which confirms our estimates of the proper motion errors from
SUPERBLINK. The proper motion errors quoted in the USNO-B1.0 catalog are
often much larger than $0.01\arcsec$ yr$^{-1}$, and the ``wings''
of the distribution in Figure 2, extending beyond $\pm0.01\arcsec$
yr$^{-1}$, correspond to those stars with larger proper motion errors
in the USNO-B1.0 catalog. There also appears to be a systematic difference
of 0.004$\arcsec$ yr$^{-1}$ in the proper motions measured in
declination by SUPERBLINK and those listed in the USNO-B1.0. The source
of this systematic difference is unclear at this point, but in any
case it is smaller than our adopted error of $\pm$0.01$\arcsec$
yr$^{-1}$ for the relative proper motions measured by SUPERBLINK.


\section{Comparison with the Luyten proper motion catalogs}

We have cross-correlated the list of high proper motion stars
identified with SUPERBLINK with a list of objects comprising all stars
listed in the two main Luyten catalogs (LHS, NLTT). Overall, 1593 of
the 1797 stars recovered by SUPERBLINK were matched to stars listed in
the Luyten catalogs. This left 204 stars not listed in the LHS or NLTT
catalog. Some 138 of those stars were found at low galactic latitudes
($|b|<25.0$) and have already been presented in Paper I. The remaining
66 new high proper motion stars, found at higher galactic latitudes,
are presented in \S4 below.

Our software also recorded 111 more stars, which we found to have
proper motions in the range $0.300<\mu<0.500\arcsec$ yr$^{-1}$, but
which were matches to stars listed in the Luyten catalogs as having
$\mu>0.500\arcsec$ yr$^{-1}$. Likewise, 56 of the stars found by
SUPERBLINK with $\mu>0.500\arcsec$ yr$^{-1}$ were listed in the LHS or
NLTT catalog as stars with proper motions $0.420<\mu<0.500\arcsec$
yr$^{-1}$. This expected blurring near the survey cutoff is the result
of proper motion errors in both Luyten's and in our own survey.

On the other hand, there are a number of northern stars listed in the
Luyten catalogs as having $\mu>0.5\arcsec$ yr$^{-1}$ that SUPERBLINK
did not recover. These include 33 stars with proper motions larger
than the upper limit of our survey ($\mu>2.0\arcsec$ yr$^{-1}$), which
our code could not possibly have recovered and are thus not considered
true ``misses''. Another 10 Luyten stars were in subfields rejected by
SUPERBLINK (see \S2). There are another 19 stars that are listed
individually in the Luyten catalog but are actually closer companions
to brighter stars, and are unresolved on the POSS plates. In all
cases, our software either did identify at least the primary star
correctly, or the unresolved pair was identified as a single moving
object. We do not now consider these as true ``misses'' either since
at least the stellar system was correctly identified. In our Paper I
analysis, however, we happened to count these stars as true misses,
which at the time gave the impression that SUPERBLINK somehow had
problems identifying moving stars in the $11<r<13$ magnitude range; this
is not the case. It is a fact, however, that SUPERBLINK systematically
misses all stars brighter than $r=8$, and misses a significant
fraction of stars in the magnitude range $9<r<11$. On the DSS images,
these very bright stars appear as very saturated, extended objects
with diffraction spikes and halos. These are not processed very
efficiently by the code, and most are simply rejected. Thus, there
were 245 very bright Luyten stars ($r<11$) that were missed by our
code. 

Finally, there were 11 faint ($12.7<r<19.2$) Luyten stars that
SUPERBLINK did not recover. We were completely unable to find 4 of
these stars (LHS1657, LHS1986, NLTT22764, NLTT11999) even by direct
visual inspection of the DSS images, and we suspect these might be
bogus. The other 7 faint high proper motion stars missed by SUPERBLINK
were found by visual inspection of the DSS images. The software missed
those for reasons that are not entirely clear, but all those objects
happen to be in the proximity of a brighter star on at least one of
the POSS images, which might explain why the software had trouble
locating them.

The new, updated distribution of northern stars with
$0.5<\mu<2.0\arcsec$ yr$^{-1}$ as a function of magnitude is shown in
Figure 3. It is compared to the old distribution based on the LHS
catalog. The peak of the distribution has shifted very little towards
fainter magnitudes and has stayed around $r=14$. While there are now
significantly more faint high proper motion stars known, there still
appear to be a sharp drop beyond $r=18$. The high success rate of
SUPERBLINK for the identification of $r=18$ high proper motion stars
strongly suggests that this feature is real, and that if any $r=19$
stars have been missed by SUPERBLINK, there should be very few of them. 

We are now in a position to directly determine the completeness of the
LHS catalog for the northern sky. Figure 4 plots the completeness of
the LHS catalog as a function of magnitude, in the assumption that the
census of high proper motion stars in the northern sky is now
complete. The figure shows that the LHS catalog was complete down to
r=12. Then the completeness dropped steadily, to fall below 50\% for
stars with r=18. Overall, the LHS catalog was about 90\% complete for
stars down to r=19, as was correctly estimated by \citet{MFLCHR00}. At
this point, however, the question of the completeness of the LHS
catalog becomes of purely historical interest, since our proper motion
survey now clearly supersedes Luyten's for the northern sky.


\section{Estimated completeness of our survey}

Our recovery rate of Luyten stars as a function of magnitude is
plotted in Figure 5. Overall, the success rate for the recovery of
Luyten stars exceeds $99.5\%$ in the magnitude range $11<r<19$.
In high galactic latitude fields, stars appear as isolated objects most
of the time, and are picked out very easily by the code. The fact that
the very few Luyten stars that SUPERBLINK missed were are in the
vicinity of brighter objects, suggest that proximity to a brighter
star must indeed be responsible for most of the non-detections. For a
star to be missed by SUPERBLINK, it has to be blended with a brighter
object in both the first and the second epoch. The chance for this to
occur depends on the local stellar density in the field, which means
that most of the stars missed by SUPERBLINK must reside in low
galactic latitude fields. The chance of a non-detection also depends
on the magnitude of the star, since it is easier for a fainter star to
hide in a crowded field than for a very bright star. For similar
reasons, the Luyten catalog was significantly more incomplete at low
galactic latitudes. 

At high galactic latitude, the probability of a chance alignment with
a brighter star at {\em both} the POSS-I and POSS-II images is very
small ($<1\%$). Hence we are confident that SUPERBLINK probably
detected $99\%$ of the high proper motion stars down to at least
$r=18$. Our main concern is with the crowded, low galactic latitude
fields, where chance alignments are much more common, and where the
probability of a non-detection is much larger.

A simple way to test the effects of crowding on the detection of high
proper motion stars is to compare the number of stars detected at high
and low galactic latitudes. A sample of high proper motion stars
selected above a fixed threshold (in our case $\mu>0.5\arcsec$
yr$^{-1}$) is not expected to be distributed uniformly across the
sky. Disk stars toward the apex and antapex tend to have smaller
proper motions than those lying in a direction perpendicular to the
Sun's motion. Since the apex is at $b\simeq23^{\circ}$, we do expect
to find fewer high proper motion, disk stars at low galactic
latitudes. Halo stars also have an asymmetric motion on the sky, and
those halo stars in the direction of galactic rotation and
antirotation are expected to show smaller proper motions. Again, this
results in a smaller number of stars, above the proper motion cutoff,
that will be found at low galactic latitudes. On the other hand, if
$N_{b>25}(r)$ is the number of stars of magnitude $r$ found at
$|b|>25$, and if $N_{b<25}(r)$ is the number of stars of magnitude $r$
found at $|b|>25$, then we expect the ratio $N_{b<25}(r)/N_{b>25}(r)$ to
be approximately constant. Since the detection of stars in crowded
fields depends on the magnitude of the star, incompleteness will show
up as a deficit of fainter stars in low galactic latitude fields.

This test is illustrated in Figure 6. The expected ratio value of the
constant $N_{b<25}(r)/N_{b>25}(r)$ is determined using the brighter
stars ($10<r<14$). For both the Luyten catalog and our survey, this
ratio is $N_{b<25}(r)/N_{b>25}(r)\simeq0.58$. The ratio is smaller
than what one would expect for a {\em uniform} distribution of stars
over the sky (0.732), which shows that our concern about expecting
fewer stars with $\mu>0.5\arcsec$ yr$^{-1}$ at low galactic latitudes
was justified. What Figure 6 reveals, however, is the significant
incompleteness of the Luyten catalogs for stars $r=15$ and
fainter. The error bars denote Poisson statistics errors. The value of
the ratio provides a reasonable estimate of the completeness at low
galactic latitude, provided that the completeness at high galactic
latitude is high. Hence, the larger ratio in the last magnitude bin
$r=19$ in Figure 6 is not significant since the census is probably
incomplete at large galactic latitudes. However, both the Luyten
survey and our own survey are significantly complete at high galactic
latitudes down to $r=18$, and hence Figure 6 provides a reasonable
means to estimate the completeness at low galactic latitudes.

Luyten's census appears to be only 50\% complete at $|b|<25$ for stars
between $r=15$ and $r=18$. Our survey, on the other hand, seems to
have recovered most of the missing stars, and appears to be
significantly complete all the way down to $r=17$. There is marginal
evidence for some missing stars at $r=15$, but the fact that the ratio
at $r=16$ and $r=17$ is close to the expected value suggests that this
is probably a statistical fluke. On the other hand, there is a
significant lack of stars in the $r=18$ magnitude bin. Assuming that
our survey is $99\%$ complete at high galactic latitudes down to
$r=18$, then we are possibly still missing $\approx40\%$ of $r=18$
stars at low galactic latitudes, which is about 9 stars.


\section{New high proper motion stars discovered by SUPERBLINK}

\subsection{Stars reported elsewhere in the literature.}

We identified 66 new high proper motion stars with $\mu>0.5\arcsec$
yr$^{-1}$ that are not listed in the Luyten catalogs. Upon
verification, we found that 9 of these had been previously reported in
the literature. These stars are listed in Table 1. Here is a short
description of how these stars were previously discovered.

\subsubsection{LSR1000+3155}

This star is the faint companion to the bright nearby G star HD86728
(GJ376), discovered by \citet{Getal00a} from 2MASS data. The companion
is an old, metal-rich M6.5 red dwarf. The system is at a distance of
14.9 pc.

\subsubsection{LSR1311+2923}

This is the star GSC2U J131147.2+292348, identified in a proper
motion survey for cool white dwarfs near the north galactic pole
\citep{CHSSLMP02}. It is a peculiar, carbon rich DQ white dwarf.
\citet{CHSSLMP02} computed the proper motion of the star using 6
photographic plates obtained at different epochs, and found a proper
motion 0.477$\pm$0.005$\arcsec$ yr$^{-1}$. This is significantly
smaller than our 0.505$\arcsec$ yr$^{-1}$, and requires further
investigation. While \citet{CHSSLMP02} cites calculated proper motions
$[\mu_{RA},\mu_{decl}]=[-0.382\pm0.002,0.286\pm0.005]\arcsec$
yr$^{-1}$, SUPERBLINK finds
$[\mu_{RA},\mu_{decl}]=[-0.374\pm0.010,0.340\pm0.010]\arcsec$
yr$^{-1}$, which shows that the discrepancy arises only in the
determination of $\mu_{decl}$. For comparison, we looked for the proper
motion of that star in the USNO-B1.0 catalog. The USNO-B1.0 quotes
this star as having proper motion components
$[\mu_{RA},\mu_{decl}]=[-0.380\pm0.004,0.332\pm0.003]\arcsec$ yr$^{-1}$
which fall within the range of values estimated by SUPERBLINK. We
therefore suspect that it is the value of $\mu_{decl}$ given by
\citet{CHSSLMP02} that is probably in error.

\subsubsection{LSR1403+3007}

This is the star 2MASSW J1403223+300755, discovered by \citet{Getal00b}
in an survey for very cool objects based on large infrared to optical
colors. The star is an M8.5 dwarf at a probable distance
d$\simeq$21.4pc.

\subsubsection{LSR1425+7102}

This star was discovered by us with our SUPERBLINK code, and was
presented in a previous paper \citep{LSR03a}. It is the first known
representative of a type of ultra-cool subdwarf (spectral type
sdM8.0).

\subsubsection{LSR1524+2925}

This is the cool star 2MASS J1524248+292535, identified in the Two
Micron All Sky Survey database from its large infrared to optical
color \citep{RKLGCM02}. The star is a nearby M7.5 dwarf, at a distance
of about 15pc.

\subsubsection{LSR1530+5608AB}

This common proper motion binary system is composed of the stars
J153055.62+560856.2 and J153055.62+560856.4. These stars were first
identified as a proper motion pair by \citet{MFLCHR00}, in a survey
for faint high proper motion stars using extra POSS II
plates. Spectroscopy shows both stars to be extreme
subdwarfs. Although \citet{MFLCHR00} do not specify the spectral
subtypes of the stars, an examination of their published Keck II
spectrum suggest that LSR1530+5608A (=J153055.62+560856.2) is an
esdM3.5, while LSR1530+5608B (=J153055.62+560856.4) is an esdM4.0.
Following the spectral type to absolute magnitude relationships
defined in \citet{LRS03b}, this places the system at a distance of
about 100pc, and suggests a transverse velocity $V_t\approx240$ km
s$^{-1}$, consistent with the pair being a member of the
Galactic halo.

\subsubsection{LSR1615+3151}

Although we found a reference to this star in the literature, it is
here identified as a high proper motion object for the first
time. This is the star CLS 100, identified as an M dwarf by
\citet{SP88}, in an objective-prism survey. Very little information
exists on this object. The absence of this star from the Luyten
catalogs comes as a surprise, since this star is relatively bright
(r=14) and does not lie in an especially crowded field.

\subsection{Stars reported for the first time.}

We present in Table 2 a set of 57 new stars with proper motion larger
than 0.5$\arcsec$ yr$^{-1}$. A search on Simbad confirms that none of
these stars has ever been mentioned in the literature before. Finder
charts for all the objects can be found in the Appendix. More than
half of the new high proper motion stars are relatively faint
($17<r<19$) objects. Such faint stars were relatively difficult to find
with Luyten's methods, which explains why they were missed.

On the other hand, our list of new high proper motion stars does
includes 24 objects with photographic red magnitude r=16 or
brighter. Such bright objects should be relatively easy to find,
which raises questions as to how they were missed by Luyten. One
notices that, while LSR0848+0856 has a red magnitude $r=13.6$, it
stands very close to the 9th magnitude star BD+09$^{\circ}$2058 in the
POSS-I image (it actually falls on one of the diffraction
spikes). This might explain why it was missed by Luyten. The
image of the $r=14.8$ star LSR0959+1739 is partially merged with that
of another bright star on the POSS-I plate, which also makes its
identification difficult; merged images of two or more stars tend
to be identified as ``non-stellar'' by plate measurement
scanners. LSR0150+0900 is also ``merged'' with another bright star on
the POSS-I image, as are LSR1122+4809, LSR1302+3340, LSR1442+3255,
LSR1604+0904, LSR1746+6953, and LSR2204+1505, while LSR1033+2559 comes
very close to being merged with another star. LSR1421+4952, on the
other hand, is partially merged with the image of a background
galaxy. Since Luyten used the POSS-I survey as his first epoch, it
thus clearly appears that proximity to a bright object was the main
source of incompleteness for bright, high proper motion stars in the
Luyten survey. Close examination also reveals that LSR0923+1817 and
LSR1340+1902, while showing up as well-isolated objects in both the
POSS-I and POSS-II images, must both have been merged with another
bright star in the 1960s, at the time when Luyten obtained second
epoch images for his proper motion survey. Our SUPERBLINK software was
specifically designed in part to address this particular problem of
``merged'' stellar images, and our results now indicate that we have
successfully achieved our goal.

The absence of LSR0939+7821, LSR1120+1953, LSR1254+7002, LSR1359+2635,
and LSR1602+0131 from the Luyten catalogs is more difficult to
explain. All these stars are reasonably bright, and make for very
obvious proper motion identification. These are true ``misses'', but
there are so few of them overall that we can only emphasize the
impressively high success rate that Luyten managed to achieve at high
galactic latitudes for stars down to r=16.


\section{Reduced proper motion diagrams and estimated stellar class}

We have constructed reduced proper motion diagrams for all the stars
recovered in our survey, by combining the USNO-B1.0 and 2MASS
magnitudes with the proper motions measured with SUPERBLINK.
The reduced proper motion in the $r$ band is defined as:
\begin{displaymath}
H_{r} \equiv r + 5 + 5 \log{\mu} = M_r + 5 \log{v_t} \-- 3.38 ,
\end{displaymath}
where $M_r$ is the absolute magnitude, and $v_t$ the transverse
velocity in km s$^{-1}$. Reduced proper motion diagrams are made by
plotting the reduced proper motion against a color, such as $H_r$
versus $b\--i$. The resulting plot is akin to an HR diagram but with an
extra scatter of the points due to the $5 \log{v_t}$ term. White
dwarfs are well separated from the red dwarfs on a reduced proper motion
diagram, just as they would be on an HR diagram. The beneficial effect
of the $5 \log{v_t}$ term is to separate old disk and halo
subdwarfs, which generally have larger $v_t$, from the young disk
dwarfs. Hence reduced proper motion diagrams are an easy way to
discriminate white dwarfs, halo subdwarfs, and disk dwarfs, in a
sample of stars for which we only have proper motions and magnitudes.

We adopt two different reduced proper motion diagrams, each of which
depends on the availability of either complete $bri$ magnitudes from
the USNO-B1.0 or $K_s$ magnitudes from 2MASS. The ($H_r$,$r\--K_s$) diagram
was investigated by \citet{SG02}, and was found to be relatively
efficient in discriminating between high proper motion dwarfs,
subdwarfs, and white dwarf stars. To cover up for those stars too
faint to show in the 2MASS survey, we also introduce here the
($H_r$,$b\--i$) diagram, which is based solely on the POSS
photographic magnitudes, obtained from the USNO-B1.0 catalog. The two
reduced proper motion diagrams are somehow complementary. Faint blue
stars, like the white dwarfs, often fail to show up on 2MASS images,
but we can almost always obtain $b\--i$ for them. On the other hand,
very red faint stars, almost invariably show up in 2MASS (in available
areas) and $r\--K_s$ is readily obtained, but these stars often do not
record on the POSS-II $b$ plates, and so are lacking $b\--i$ colors.

We plot both the ($H_r$,$r\--K_s$) and ($H_r$,$b\--i$) diagrams in Figure
7. The left panels show the diagrams for the recovered LHS stars, for
which the appropriate magnitudes are available. The right panels show
the diagrams for the new high proper motion stars found by SUPERBLINK.
Dashed lines separate the loci of the three general types of objects
that are discriminated in the reduced proper motion diagrams. The
regions are labeled ``wd'' for the white dwarfs, ``sd'' for the
subdwarfs or halo dwarfs, and ``d'' for the disk dwarfs. The
boundaries between the three regions are somewhat arbitrary, as there
always is a significant scatter of data points in reduced proper
motion diagrams. For the ($H_r$,$r\--K_s$) diagram, we use the limits
defined by \citet{SG02}. To guide the tracing of the boundaries in the
($H_r$,$b\--i$) diagram, we have separately plotted on this diagram the
stars that fell in each of the regions in the ($H_r$,$r\--K_s$) diagram,
and for which both $r\--K_s$ and $b\--i$ was available.

We find the ($H_r$,$b\--i$) diagram to be just as efficient as
($H_r$,$r\--K_s$) in discriminating between white dwarfs and main
sequence stars. On the other hand, the red dwarfs and subdwarfs are
not as clearly separated in ($H_r$,$b\--i$) as they are in
($H_r$,$r\--K_s$). There appears to be a larger scatter in $H_r$
for a given value of $b\--i$ than there is for a given value of
($r\--K_s$). This is possibly due to the larger instrumental errors in
$b\--i$, but may also reflect an actual intrinsic scatter in the
$b\--i$ colors of red dwarfs and subdwarfs of a given luminosity,
which could be the result e.g. of atmospheric activity in some of the
stars. 

We have determined stellar classes for all but one of the new high
proper motion stars listed in Table 2. These stars are featured in at
least one of the reduced proper motion diagrams. Several stars show up
in both, and their position on the ($H_r$,$r\--K_s$) diagram was
generally consistent with their position on the ($H_r$,$b\--i$)
diagram. However there were a number of cases in which the positions were
conflicting, and for those we gave precedence to the ($H_r$,$r\--K_s$)
diagram, following our discussion above. Individual stellar classes
are included in Table 2. One should keep in mind that the classes are
only indicative, and that spectroscopy is required to confirm the
class of a object. Stellar classes determined from the reduced proper
motion diagram are however useful as a guide to follow-up
observations.

Figure 7 shows that a particularly significant result of our new
survey is the addition of a significant number of stars with very
large reduced proper motion. The number of known stars with $H_r>21$
has been nearly doubled. Most of these objects are faint white dwarfs
and subdwarfs. Follow-up observations of these stars could lead to
significant advances in our knowledge of old white dwarfs and low-mass
subdwarfs.


\section{Conclusions}

Our semi-automated survey for high proper motion in the northern sky,
using the SUPERBLINK software, has been expanded and now covers over
97\% of the sky north of $\delta=0.0^{\circ}$. In Paper I, we had
reported the discovery of 138 new northern ($\delta=0.0^{\circ}$)
stars with proper motions $0.5<\mu<2.0=\arcsec$ yr$^{-1}$ at low
galactic latitudes ($|b|<25$). To this, we now add 57 new high proper
motion stars at high galactic latitudes ($|b|>25$). Our survey now
includes 1797 northern stars with $0.5<\mu<2.0=\arcsec$ yr$^{-1}$, all
in the magnitude range $7<r<20$. Most objects are rediscoveries of
stars listed in the LHS and NLTT catalogs. Our extremely high recovery
rate ($>99.5\%$) of LHS and NLTT stars in the range $11<r<19$ suggest
that our new survey has a very high completeness level in that
magnitude range.

The SUPERBLINK software did not recover any star fainter than $r=20$
which is the magnitude limit of the POSS-I plates, used as our first
epoch. The efficiency of SUPERBLINK was also shown to be limited for
brighter stars ($r<11$), and nil for stars brighter than $r=7$. This
however, is of little consequence since the vast majority of $r<11$
stars on the sky have already had their proper motions
measured with the HIPPARCOS satellite. A combination of the bright
high proper motion stars listed in the Tycho-2 catalog \citep{H00}
with our own list of objects should yield a very nearly complete, and
most accurate list of northern high proper motion stars with
$0.5<\mu<2.0=\arcsec$ yr$^{-1}$ down to magnitude $r=19$.

We have directly determined, once and for all, the completeness of the
LHS catalogue for the northern sky. Overall, Luyten's census of stars
with proper motions in the range $0.5<\mu<2.0=\arcsec$ yr$^{-1}$ was a
little less than 90\% complete for stars down to $r=19$. However, his
survey was apparently complete down to $r=13$, but significantly
incomplete at fainter magnitudes. We hope to have now settled the
debate about the completeness of the LHS catalog. Our survey now
supersedes the Luyten proper motion surveys, at least for the northern
sky.

Reduced proper motion diagrams show that the newly discovered high
proper motion stars are a mix of faint and probably nearby disk
dwarfs, faint halo dwarfs or subdwarfs, and faint white
dwarfs. While overall our survey added only $10\%$ new objects
to the census of stars with proper motions $\mu>0.5\arcsec$ yr$^{-1}$,
this contribution is very significant because most of the new stars
are faint, low luminosity objects. In particular, the census of faint
high proper motion white dwarfs and subdwarfs has been nearly
doubled. We are currently working on a spectroscopic follow-up program
for the high proper motion stars found in our survey. Spectral
classification for most of the stars reported in this paper will be
presented in L\'epine {\it et al.} (2003, in preparation).


\acknowledgments

{\bf Acknowledgments}

This research program has been supported by NSF grant AST-0087313 at
the American Museum of Natural History, as part of the NSTARS program.
This work has been made possible through the use of the Digitized Sky
Surveys. The Digitized Sky Surveys were produced at the Space
Telescope Science Institute under U.S. Government grant NAG
W-2166. The images of these surveys are based on photographic data
obtained using the Oschin Schmidt Telescope on Palomar Mountain and
the UK Schmidt Telescope. The plates were processed into the present
compressed digital form with the permission of these institutions. The
National Geographic Society - Palomar Observatory Sky Survey (POSS I)
was made by the California Institute of Technology with grants from
the National Geographic Society. The Second Palomar Observatory Sky
Survey (POSS II) was made by the California Institute of Technology
with funds from the National Science Foundation, the National
Geographic Society, the Sloan Foundation, the Samuel Oschin
Foundation, and the Eastman Kodak Corporation. The Oschin Schmidt
Telescope is operated by the California Institute of Technology and
Palomar Observatory. The UK Schmidt Telescope was operated by the
Royal Observatory Edinburgh, with funding from the UK Science and
Engineering Research Council (later the UK Particle Physics and
Astronomy Research Council), until 1988 June, and thereafter by the
Anglo-Australian Observatory. The blue plates of the southern Sky
Atlas and its Equatorial Extension (together known as the SERC-J), as
well as the Equatorial Red (ER), and the Second Epoch [red] Survey
(SES) were all taken with the UK Schmidt. 

The data mining required for this work has been made possible with the
use of the SIMBAD astronomical database and VIZIER astronomical
catalogs service, both maintained and operated by the Centre de
Donn\'ees Astronomiques de Strasbourg (http://cdsweb.u-strasbg.fr/).

\newpage

\appendix

\section{Finder charts}

Finder charts of high proper motion stars are generated as a
by-product of the SUPERBLINK software. We present in Figures 8a-8d
finder charts for all the new, high proper motion stars presented in
this paper and listed in Table1. All the charts consist of pairs of
$4.25\arcmin\times4.25\arcmin$ images showing the local field at two
different epochs. The name of the star is indicated in the center just
below the chart, and corresponds to the name given in Table 1. To the
left is the POSS-I field, with the epoch of the field noted in the
lower left corner (typically in the 1953-1955 range). To the right is
the modified POSS-II field which has been shifted, rotated, and
degraded in such a way that it matches the quality and aspect of the
POSS-I image. The epoch of the POSS-II field is noted on the lower
right corner. High proper motion stars are identified with circles
centered on their positions at the epoch on the plate.

The charts are oriented in the local X-Y coordinate system of the
POSS-I image; the POSS-II image has been remapped on the POSS-I
grid. This means that north is generally up and east left, but the
fields might appear rotated by a small angle for high declination
objects. Sometimes a part of the field is missing: this is an artifact
of the code. SUPERBLINK works on $17\arcmin\times17\arcmin$ DSS
subfields. If a high proper motion star is identified near the edge of
that subfield, the output chart will be truncated. Similar finder
charts are automatically generated for every single object identified
by the code, and we are currently building a large electronic catalog
of two-epoch finder charts.


\newpage

\begin{deluxetable}{llrrrrrrrrrr}
\tabletypesize{\scriptsize}
\tablecolumns{13}
\tablewidth{0pt}
\tablecaption{LSR stars not in the Luyten catalogs, but reported elsewhere in the literature}
 \tablehead{
\colhead{Star} & 
\colhead{$\alpha$(J2000)} &
\colhead{$\delta$(J2000)} &
\colhead{$\mu$($\arcsec$ yr$^{-1}$)} & 
\colhead{$pma$($^{\circ}$)} & 
\colhead{$b$\tablenotemark{a}} & 
\colhead{$r$} &
\colhead{$i$} & 
\colhead{$J$\tablenotemark{b}} &
\colhead{$H$} &
\colhead{$K_s$} &
\colhead{ref\tablenotemark{c}}
}
\startdata 
LSR1000+3155 & 10 00 50.70& +31 55 49.4& 0.511& 231.9& 17.0& 14.9& 12.0& 10.26&  9.64&  9.27& G00a\\ 
LSR1311+2923 & 13 11 46.92& +29 23 52.5& 0.505& 312.2& 19.1& 17.7& 17.9&   .  &   .  &   .  & C02 \\
LSR1403+3007 & 14 03 22.20& +30 07 56.7& 0.796& 273.7& 20.9& 18.7& 15.2& 12.68& 12.00& 11.60& G00b\\
LSR1425+7102 & 14 25 04.81& +71 02 10.4& 0.635& 254.7& 20.7& 18.6& 16.1& 14.77& 14.40& 14.33& L03 \\
LSR1524+2925 & 15 24 24.73& +29 25 31.6& 0.626& 184.8& 20.3& 16.6& 13.3& 11.21& 10.53& 10.15& R02 \\
LSR1530+5608A& 15 30 50.32& +56 08 43.5& 0.545& 298.9& 19.4& 17.5& 16.0& 15.19& 14.57& 14.32& M00 \\
LSR1530+5608B& 15 30 49.99& +56 08 50.1& 0.515& 298.5& 20.1& 17.8& 16.3& 15.47& 14.93& 14.76& M00 \\
LSR1615+3151 & 16 15 09.98& +31 51 47.1& 0.756& 179.4& 16.3& 14.1& 11.4& 10.34&  9.85&  9.55& S88
\enddata                              
\tablenotetext{a}{Photographic $bri$ magnitudes from the USNO-B1.0 catalog.}
\tablenotetext{b}{Infrared $JHK_s$ magnitudes from the 2MASS All-Sky
  Point Source Catalog.}
\tablenotetext{c}{Reference code: C02 = \citet{CHSSLMP02}, G00a =
\citet{Getal00a}, G00b = \citet{Getal00a}, L03 = \citet{LSR03a}, G97 =
\citet{GSIJ97}, R02 = \citet{RKLGCM02}, M00 = \citet{MFLCHR00}, S88 =
\citet{SP88}}
\end{deluxetable}                     

\begin{deluxetable}{llrrrrrrrrrr}
\tabletypesize{\scriptsize}
\tablecolumns{13}
\tablewidth{0pt}
\tablecaption{LSR stars not in the Luyten catalogs, and reported here for the first time.}
\tablehead{
\colhead{Star} & 
\colhead{$\alpha$(J2000)} &
\colhead{$\delta$(J2000)} &
\colhead{$\mu$} & 
\colhead{$pma$} & 
\colhead{$b$\tablenotemark{a}} & 
\colhead{$r$} &
\colhead{$i$} & 
\colhead{$J$\tablenotemark{b}} &
\colhead{$H$} &
\colhead{$K_s$}&
\colhead{class}\\
\colhead{} & 
\colhead{} &
\colhead{} &
\colhead{($\arcsec$ yr$^{-1}$)} & 
\colhead{($^{\circ}$)} & 
\colhead{} & 
\colhead{} &
\colhead{} & 
\colhead{} &
\colhead{} &
\colhead{} &
\colhead{}
}
\startdata 
LSR0018+2853&  00 18 06.48& +28 53 27.9& 0.534&  89.2& 20.9&  19.2&  17.7&  16.19&   15.62&   15.49 & sd \\
LSR0053+2054&  00 53 17.64& +20 54 45.4& 0.559&  92.1& 20.2&  18.3&  15.5&  13.50&   12.94&   12.69 &  d \\
LSR0101+3521&  01 01 27.32& +35 21 53.9& 0.606&  92.2& 20.1&  18.8&  16.5&  15.59&   15.22&   14.88 & sd \\
LSR0150+0900&  01 50 36.18& +09 00 22.9& 0.578& 196.6& 19.4&  16.3&  14.6&  13.61&   13.15&   12.87 & sd \\
LSR0302+0021&  03 02 28.76& +00 21 43.5& 0.504& 146.1& 20.1&  17.5&  15.8&  14.77&   14.12&   14.03 & sd \\
LSR0328+1129&  03 28 34.58& +11 29 52.4& 0.599&  64.7& 21.6&  18.9&  15.2&  12.46&   11.78&   11.33 &  d \\
LSR0804+6153&  08 04 05.84& +61 53 33.4& 0.635& 184.6& 21.3&  18.8&  15.2&  12.74&   11.93&   11.45 &  d \\
LSR0822+1700&  08 22 33.78& +17 00 20.1& 0.605& 143.5&\nodata&  18.7&  17.2&  15.72&   15.52&   15.62 & sd \\
LSR0833+3706&  08 33 03.17& +37 06 08.4& 0.625& 213.4& 19.5&  17.9&  14.8&  12.28&   11.63&   11.23 &  d \\
LSR0836+2432&  08 36 18.04& +24 32 56.3& 0.542& 154.3& 20.1&  19.0&  18.1&\nodata& \nodata& \nodata & wd \\
LSR0837+7037&  08 37 18.03& +70 37 33.8& 0.816& 173.2& 18.1&  16.4&  14.0&  12.73&   12.25&   11.97 &  d \\
LSR0848+0856&  08 48 01.16& +08 56 48.3& 0.702& 196.3& 15.4&  13.6&  12.4&  11.28&   10.73&   10.50 & sd \\
LSR0923+1817&  09 23 32.43& +18 17 05.8& 0.858& 136.8& 18.0&  16.0&  14.8&  13.80&   13.42&   13.18 & sd \\
LSR0939+7821&  09 39 10.87& +78 21 29.2& 0.987& 231.2& 18.3&  15.6&  12.8&  11.73&   11.26&   10.97 &  d \\
LSR0959+1739&  09 59 41.94& +17 39 20.4& 0.541& 162.3& 17.0&  14.8&  13.1&  12.22&   11.72&   11.50 & sd \\
LSR0959+7533&  09 59 55.31& +75 33 09.6& 0.557& 220.6&\nodata&  18.0&  14.7&  12.57&   11.98&   11.60 &  d \\
LSR1002+6108&  10 02 25.78& +61 08 58.9& 0.561& 234.8& 19.5&  18.5&  17.8&\nodata& \nodata& \nodata & wd \\
LSR1002+6349&  10 02 12.10& +63 49 26.1& 0.554& 252.0& 20.8&  18.7&  15.5&  12.90&   12.36&   11.98 &  d \\
LSR1005+3759&  10 05 49.78& +37 59 03.3& 0.517& 249.2& 19.2&  16.5&  13.1&  12.00&   11.44&   11.10 &  d \\ 
LSR1033+2559&  10 33 20.37& +25 59 22.3& 0.754& 139.5& 17.7&  15.6&  12.9&  11.80&   11.24&   10.93 &  d \\
LSR1044+1327&  10 44 53.65& +13 27 57.5& 0.527& 253.4& 20.3&  18.8&  18.6&\nodata& \nodata& \nodata & wd \\
LSR1107+4855&  11 07 31.38& +48 55 23.0& 0.729& 263.9& 20.7&  18.7&  18.0&\nodata& \nodata& \nodata & wd \\
LSR1118+0941&  11 18 14.64& +09 41 13.7& 0.508& 179.0& 20.1& (19.3)& 17.4&  15.92&   15.38&   15.25 & sd \\
LSR1119+0820&  11 19 46.47& +08 20 36.4& 0.517& 132.5& 20.6& (18.7)& 15.0&  12.77&   12.22&   11.90 &  d \\
LSR1120+1953&  11 20 27.61& +19 53 28.7& 0.846& 231.1& 17.2&  14.9&  13.2&  11.16&   10.65&   10.38 &  d \\
LSR1122+4809&  11 22 29.51& +48 09 54.6& 0.599& 268.6& 18.1&  15.9&  13.9&  12.16&   11.55&   11.30 &  d \\
LSR1143+0413&  11 43 51.38& +04 13 26.0& 0.534& 257.8& 20.2&  18.3&  14.8&  13.11&   12.51&   12.13 &  d \\
LSR1153+3414&  11 53 48.70& +34 14 16.4& 0.511& 130.2& 18.3&  16.2&  14.4&  13.29&   12.78&   12.56 & sd \\
LSR1204+3158&  12 04 14.48& +31 58 05.0& 0.571& 264.5& 20.4&  18.9& \nodata&  15.54&   14.97&   14.65 & sd \\
LSR1227+2512&  12 27 42.21& +25 12 58.8& 0.581& 289.6& 20.6&  18.6&  17.3&  14.79&   14.26&   14.12 & sd \\
LSR1254+7002&  12 54 37.92& +70 02 49.2& 0.861& 249.7&\nodata&  14.9&  13.6&  12.99&   12.48&   12.25 & sd \\
LSR1259+1956A& 12 59 45.82& +19 56 58.2& 0.501& 291.2& 20.2&  17.9&  14.8&  13.45&   12.95&   12.73 &  d \\
LSR1259+1956B& 12 59 48.36& +19 56 41.0& 0.495& 291.0& 20.3&  18.9&  15.3&  13.77&   13.28&   13.00 &  d \\
LSR1302+3339&  13 02 52.41& +33 39 59.7& 0.609& 276.8& 17.5&  15.4&  13.8&  13.73&   13.23&   13.07 & sd \\
LSR1305+1934&  13 05 36.72& +19 34 57.4& 0.558& 337.3& 19.8&  16.8&  13.8&  12.13&   11.51&   11.14 &  d \\ 
LSR1334+3303&  13 34 29.21& +33 03 04.0& 0.649& 267.3& 19.8&  17.3&  14.3&  12.50&   11.93&   11.61 &  d \\
LSR1340+1902&  13 40 40.74& +19 02 22.2& 0.912& 209.3& 19.7&  16.8&  14.5&\nodata& \nodata& \nodata &  d \\
LSR1359+2635&  13 59 35.52& +26 35 31.6& 0.628& 248.8& 18.0&  15.3&  12.5&  11.27&   10.75&   10.47 &  d \\
LSR1421+4952&  14 21 25.55& +49 52 02.5& 0.602& 264.7& 16.9&  15.3&  13.3&  13.02&   12.47&   12.30 & sd \\
LSR1442+3255&  14 42 06.56& +32 55 07.4& 0.526& 168.4& 16.7&  14.9&  14.0&  12.93&   12.39&   12.16 & sd \\ 
LSR1448+6148&  14 48 46.86& +61 48 02.5& 0.976& 170.9& 20.6&  17.9&  15.9&  14.84&   14.53&   14.01 & sd \\
LSR1523+3152&  15 23 40.67& +31 52 57.1& 0.628& 165.7& 20.5&  18.6&  18.3&\nodata& \nodata& \nodata & wd \\
LSR1537+6102&  15 37 38.77& +61 02 46.4& 0.555& 279.4& 19.1&  16.9&  15.3&  14.51&   14.11&   13.94 & sd \\
LSR1554+1639&  15 54 00.21& +16 39 50.6& 0.523& 225.4& 20.9&  18.6&  15.3&  13.11&   12.52&   12.16 &  d \\
LSR1559+7313&  15 59 34.78& +73 13 59.2& 0.501& 309.2& 18.8&  18.3&  18.6&\nodata& \nodata& \nodata & wd \\
LSR1601+5943&  16 01 12.35& +59 43 25.6& 0.502& 168.6& 19.3&  17.7&  17.8&\nodata& \nodata& \nodata & wd \\
LSR1602+0131&  16 02 41.71& +01 31 57.3& 0.581& 196.4& 18.3&  15.4&  12.4&  10.89&   10.39&   10.03 &  d \\
LSR1604+0904&  16 04 19.09& +09 04 29.6& 0.581& 273.5& 18.4&  16.3&  13.4&  12.32&   11.83&   11.56 &  d \\
LSR1641+2449&  16 41 23.86& +24 49 44.0& 0.650& 255.1& 20.8&  17.9&  16.9&  15.36&   14.98&   14.99 & sd \\
LSR1703+5910&  17 03 14.25& +59 10 48.6& 0.571& 147.8& 19.8&  17.7&  14.8&  12.81&   12.15&   11.86 &  d \\
LSR1746+6953&  17 46 49.03& +69 53 26.9& 0.538& 198.9& 17.4&  14.6&  13.9&  12.79&   12.26&   12.06 & sd \\
LSR1940+8348&  19 40 08.59& +83 48 58.3& 0.936& 243.5& 18.7&  17.5&  17.2&  16.65&   16.24&   16.96 & wd \\
LSR2204+1505&  22 04 21.50& +15 05 52.0& 0.816&  52.1& 16.8&  14.7&  12.2&  10.87&   10.31&    9.98 &  d \\
LSR2222+1221&  22 22 33.76& +12 21 43.0& 0.728&  74.9& 20.0&  16.5&  17.5&\nodata& \nodata& \nodata & sd \\
LSR2223+1602&  22 23 21.86& +16 02 09.6& 0.635& 152.2& 20.3&  18.5&  15.0&  14.10&   13.58&   13.29 &  d \\ 
LSR2312+1532&  23 12 39.75& +15 32 38.5& 0.564& 101.1& 20.6&  18.6&  15.5&  13.32&   12.73&   12.38 &  d \\ 
LSR2338+3332&  23 38 26.13& +33 32 50.6& 0.639& 165.6&\nodata& (19.1)&\nodata&\nodata& \nodata& \nodata & --
\enddata                              
\tablenotetext{a}{Photographic $bri$ magnitudes from the USNO-B1.0
catalog. Values shown in parenthesis are magnitude estimates provided
by the SUPERBLINK software in the absence of a corresponding value in
the USNO-B1.0 catalog.}
\tablenotetext{b}{Infrared $JHK_s$ magnitudes from the 2MASS All-Sky
Point Source Catalog.}
\end{deluxetable}                     

\newpage

\begin{figure}
\plotone{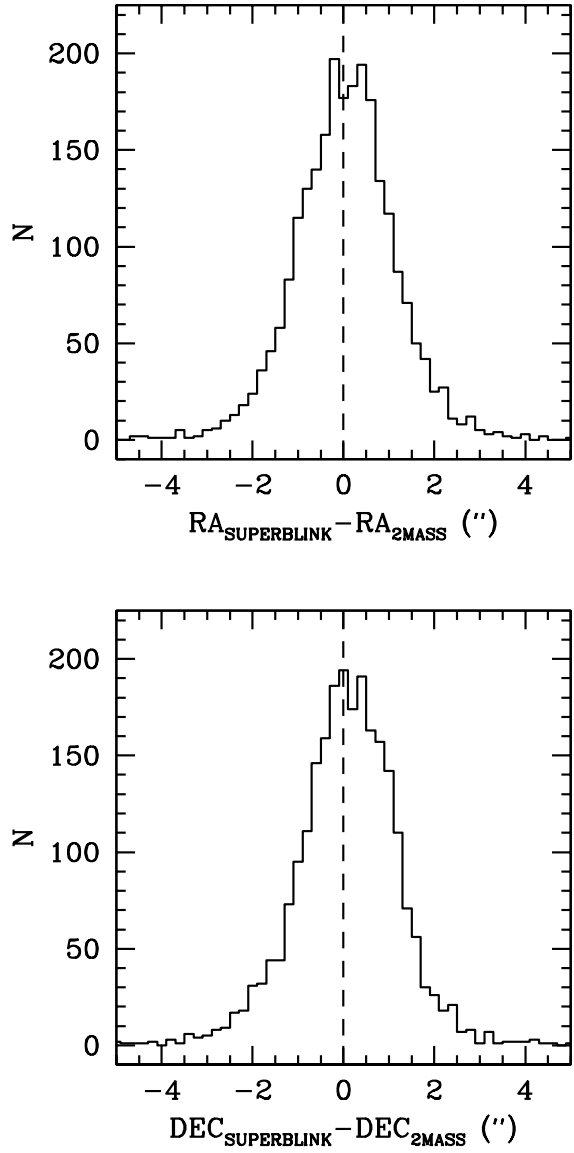}
\caption{Difference between the positions of 1754 high proper motion stars
given by our SUPERBLINK software with the positions of the same stars
in the 2MASS All-Sky Point Source Catalog.}
\end{figure}

\begin{figure}
\plotone{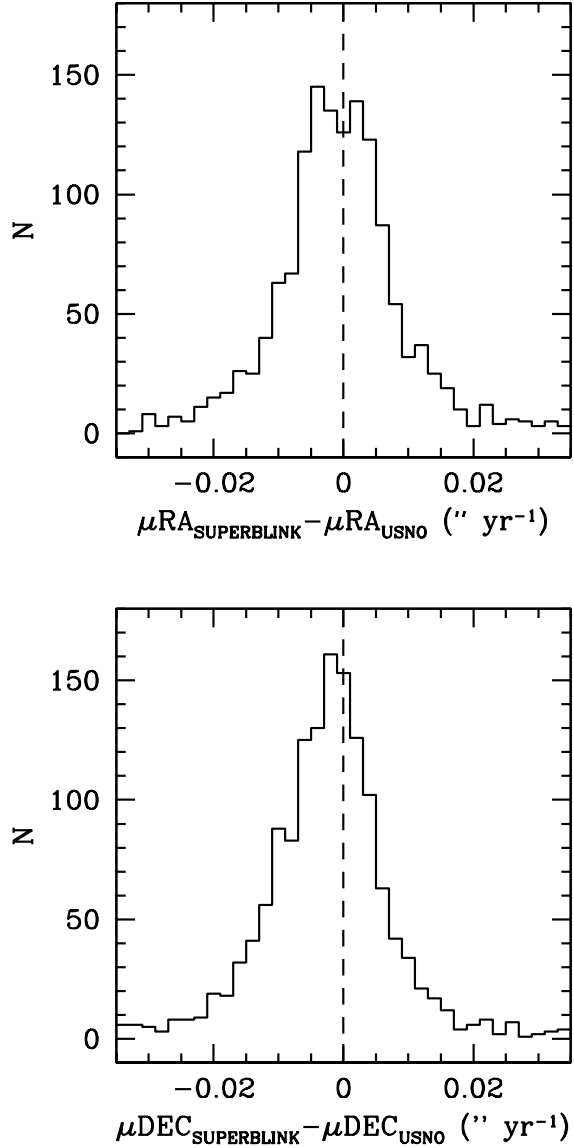}
\caption{Difference in the proper motions measured by SUPERBLINK and
those listed for the same stars recovered in the USNO-B1.0
catalog. The statistical distribution is shown separately for proper
motions in RA (top) and DEC (bottom). Proper motion errors from
SUPERBLINK are estimated to be $\pm0.01\arcsec$ yr$^{-1}$ in both RA
and DEC, consistent with the distribution shown here. Estimated proper
motions in the USNO-B1.0 catalog are generally $\pm0.005\arcsec$
yr$^{-1}$, although much larger errors exist for a significant
fraction ($\sim15\%$) of the stars, as evidenced here by the broad
``wings'' of the distribution.}
\end{figure}

\begin{figure}
\plotone{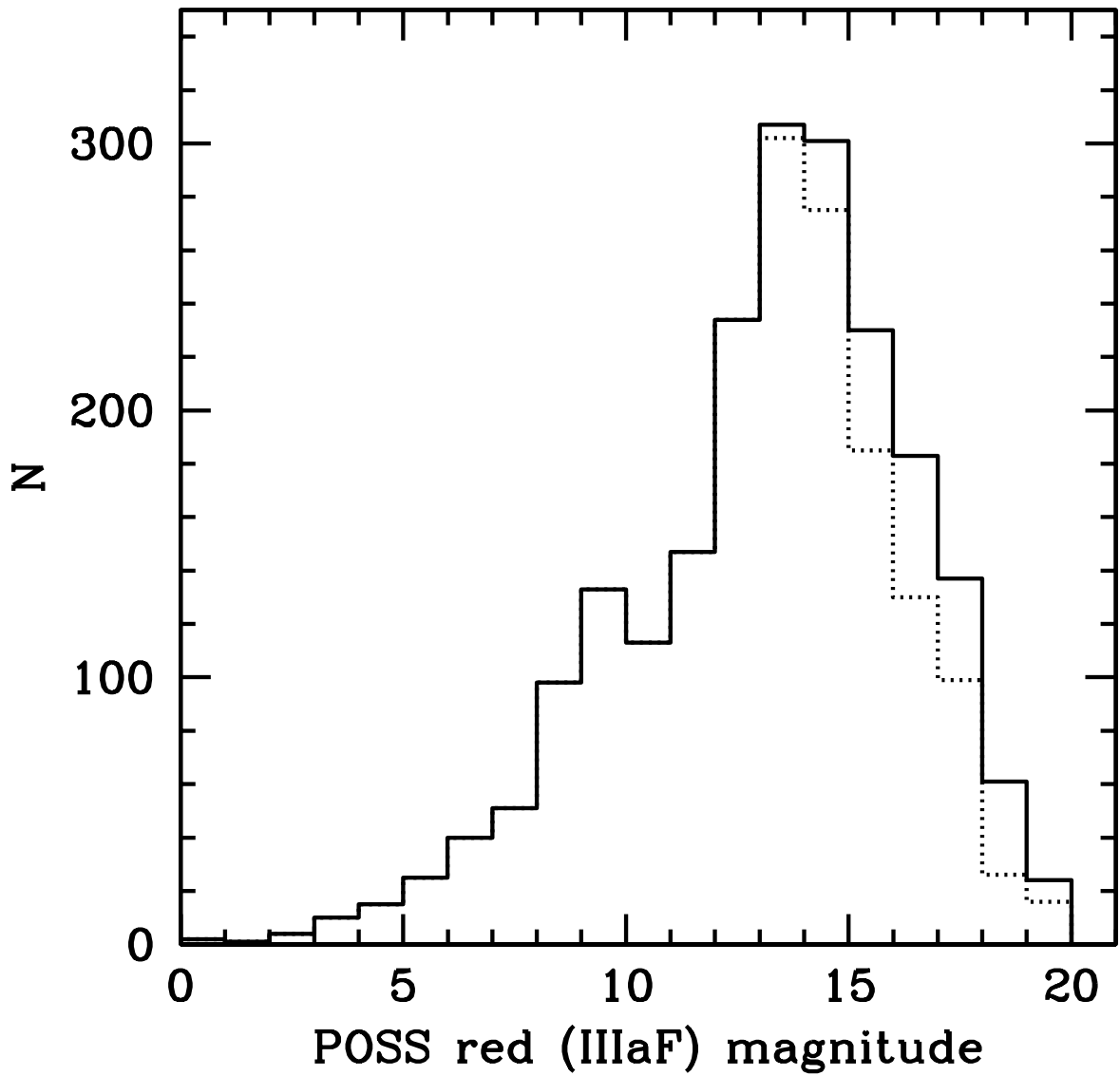}
\caption{Updated distribution of northern stars with proper motions
$0.5<\mu<2.0\arcsec$ yr$^{-1}$ as a function of the photographic red
(IIIaf) magnitude (full line). The previous distribution, based on LHS
catalog stars, is shown for comparison (dotted line).}
\end{figure}

\begin{figure}
\plotone{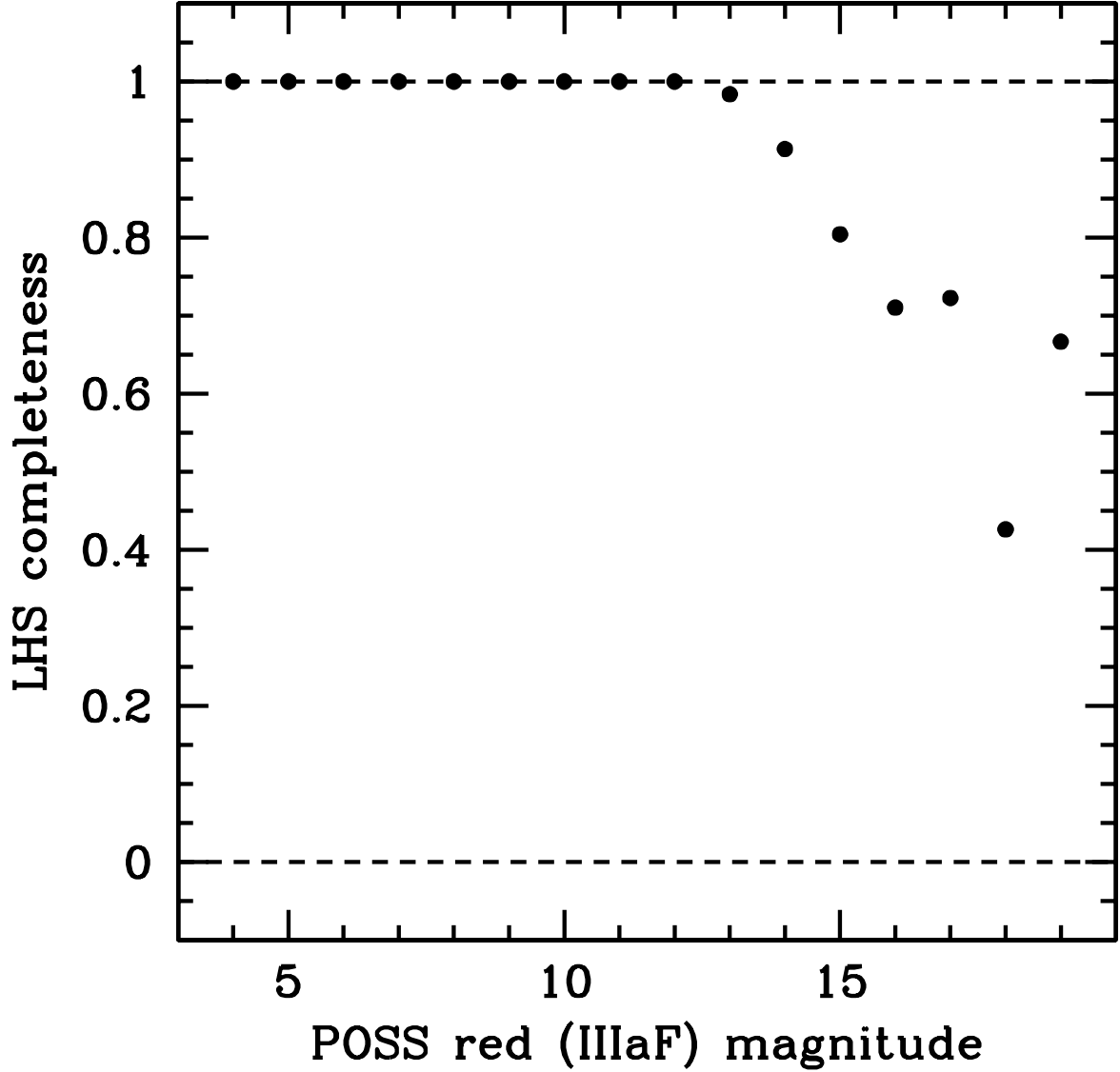}
\caption{Measured completeness of the LHS catalog in the northern sky,
as a function of photographic red (IIIaF) magnitude, for stars with
proper motion $0.5<\mu<2.0\arcsec$ yr$^{-1}$. The LHS catalog was
apparently complete for stars brighter than r=13, but its completeness
dropped steadily at fainter magnitudes, falling below 50\% at
$r=18$. Our new survey now considerably increases the completeness of
the high proper motion census in that range.}
\end{figure}

\begin{figure}
\plotone{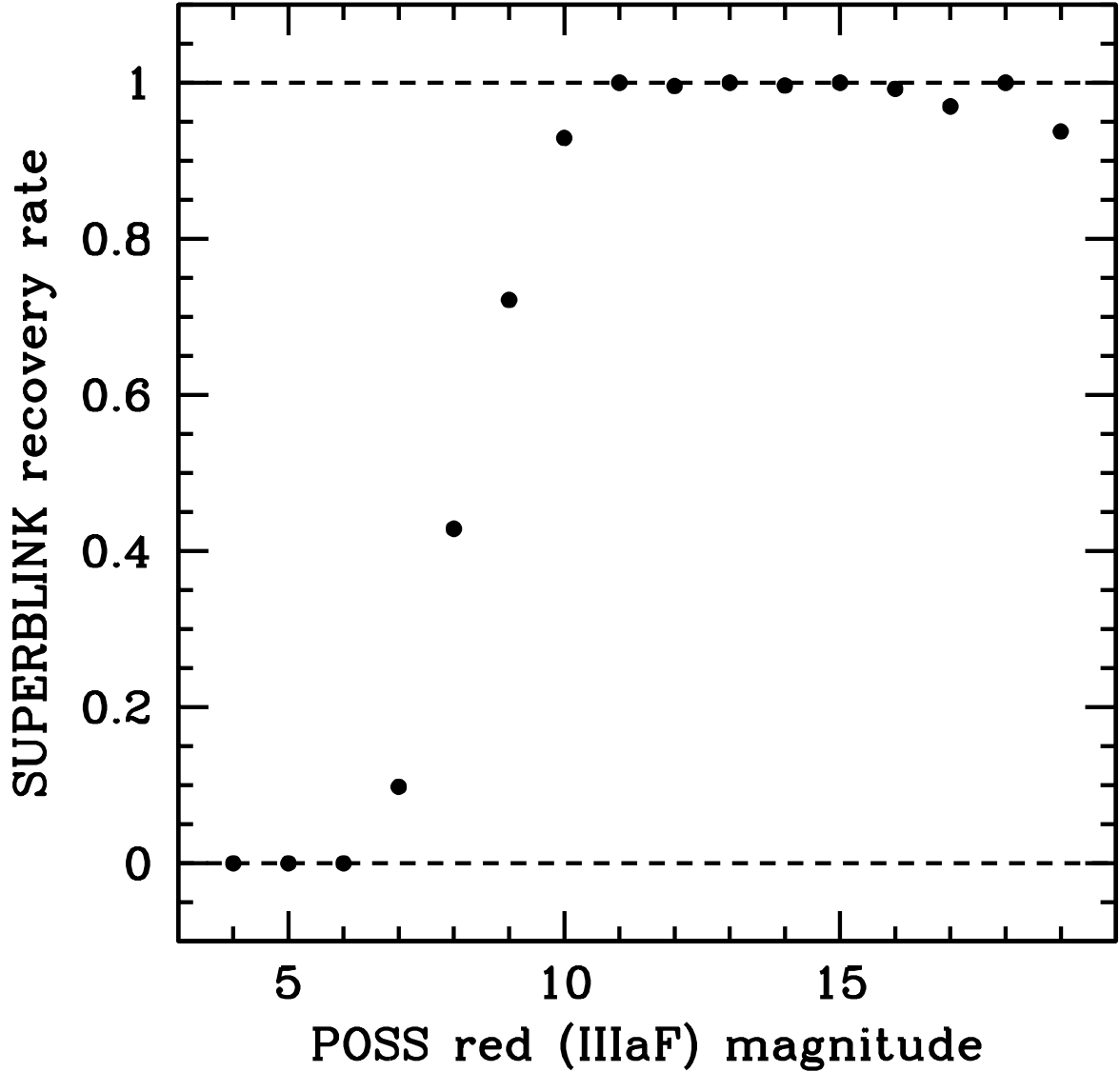}
\caption{Recovery rate of LHS stars with proper motions
($0.5<\mu<2.0\arcsec$ yr$^{-1}$) by the SUPERBLINK software, as a
function of the photographic red (IIIaF) magnitude.
While SUPERBLINK is relatively inefficient in identifying very bright
($r<10$) high proper motion stars, it is remarkably efficient for
fainter stars, down to $r=19$. Overall, more than 99.5\% of the
faint ($11<r<19$) LHS stars with proper motions $0.5<\mu<2.0\arcsec$
yr$^{-1}$, have been recovered by SUPERBLINK in the area covered by
our survey.}
\end{figure}

\begin{figure}
\plotone{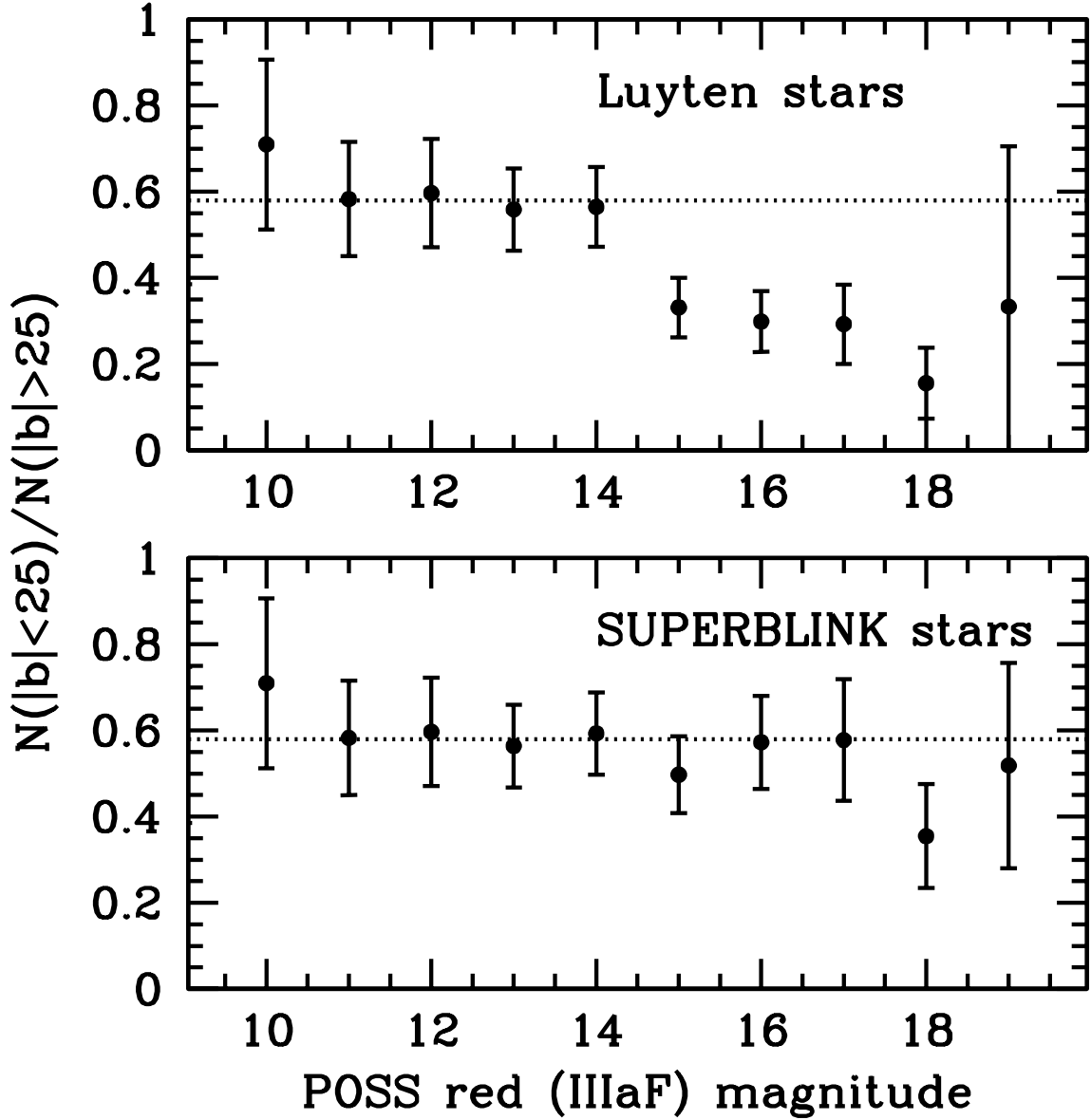}
\caption{Ratio between the number of high proper motion stars
found at low galactic latitudes ($|b|<25.0$) and the number of high
proper motion stars found at high galactic latitudes ($|b|>25.0$), for
the whole northern sky. The ratio is plotted separately for each
magnitude bin between $r=10$ and $r=19$. The ratio is expected to be
independent on the magnitude. This clearly shows the Luyten survey
(top) to be significantly incomplete at $|b|<25.0$, for stars $r=15$
or fainter. Our own survey (bottom), shows evidence for significant
incompleteness only for $r=18$ stars.}
\end{figure}

\begin{figure}
\plotone{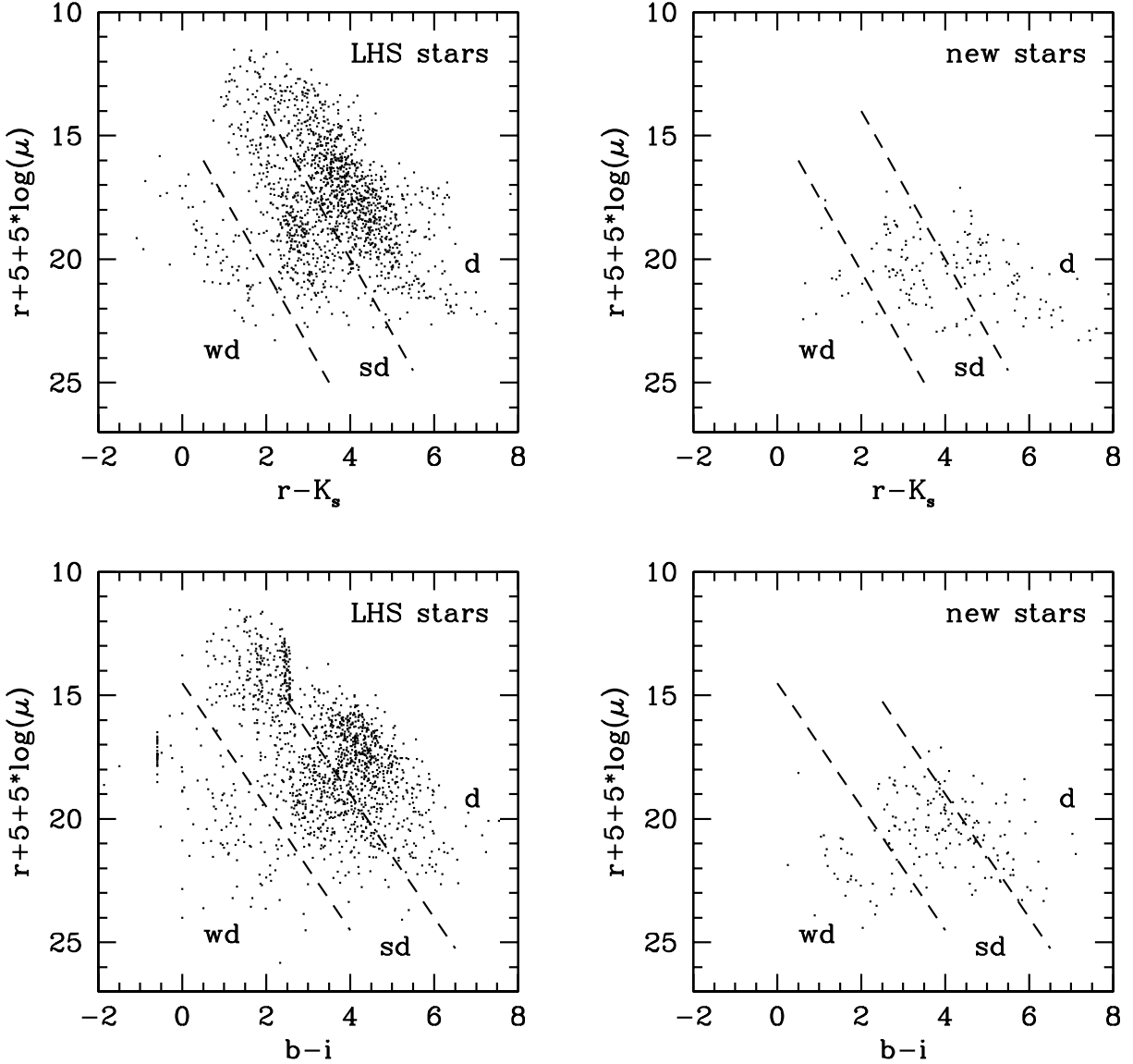}
\caption{Reduced proper motion diagram for the high proper motion
stars recovered by SUPERBLINK in high Galactic latitude fields. Left
panels: star previously listed in the LHS and/or NLTT catalog. Right
panels: new stars. Upper panels show the diagrams built with the
$r-K_s$ color term, and include only those stars which have a
counterpart in the 2MASS All-Sky Point Source Catalog. Bottom panels are
built with the $b-i$ color term, and include all stars for which both
a $b$ and $i$ magnitude was found in the USNO-B1.0 catalog. Dashed
lines delimitate the region typically occupied by dwarf disk stars (d),
subdwarf halo stars (sd), and white dwarfs (wd).}
\end{figure}

\begin{figure}
\caption{Finding charts for the new high proper motion stars
discovered in our survey, as listed in Tables 1 and 2.}
\end{figure}

\begin{figure}
\caption{Finding charts for the new high proper motion stars
discovered in our survey (continued).}
\end{figure}

\begin{figure}
\caption{Finding charts for the new high proper motion stars
discovered in our survey (continued).}
\end{figure}

\begin{figure}
\caption{Finding charts for the new high proper motion stars
discovered in our survey (continued).}
\end{figure}


\begin{thebibliography}{}

\bibitem[Bakos, Sahu, \& Nemeth(2002)]{BSN02}
Bakos, G. A., Sahu, K. C., Nemeth, P. 2002, \apjs, 141, 187

\bibitem[Carollo {\it et al.}(2002)]{CHSSLMP02}
Carollo, D., Hodgkin, S. T., Spagna, A., Smart, R. L., Lattanzi,
M. G., McLean, B. J., \& Pinfield, D. J. 2002, \aap, 393, L45

\bibitem[Gould(2003)]{G03}
Gould, A. 2003, \aj, {\it in press}

\bibitem[Gizis {\it et al.}(1997)]{GSIJ97}
Gizis, J. E., Scholz, R.-D., Irwin, M., \& Jahreiss, H. 1997, \mnras,
292, L41

\bibitem[Gizis {\it et al.}(2000a)]{Getal00a}
Gizis, J. E., Monet, D. G., Reid, I.N., Kirkpatrick, J. D., \&
Burgasser, A. J. 2000, \mnras, 311, 385

\bibitem[Gizis {\it et al.}(2000b)]{Getal00b}
Gizis, J. E., Monet, D. G., Reid, I. N., Kirkpatrick, J. D., Liebert,
J., \& Williams, R. J. 2000 \aj, 120, 1085

\bibitem[Hog {\it et al.}(2000)]{H00} 
Hog E., Fabricius C., Makarov V.V., Urban S., Corbin T.,
Wycoff G., Bastian U., Schwekendiek P., \& Wicenec A. 2000, The
Tycho-2 Catalogue of the 2.5 Million Brightest Stars, \aap 355, 27
({\it CDS-ViZier catalog number I/259})

\bibitem[L\'epine Shara, \& Rich(2002)]{LSR02b}
L\'epine, S., Shara, M. M., \& Rich, R. M. 2002, \aj, 124, 1190

\bibitem[L\'epine {\it et al.}(2002)]{LRNCS02c}
L\'epine, S., Rich, R. M., Neill, J. D., Caulet, A., \& Shara,
M. M. 2002, \apjl, 581, L47

\bibitem[L\'epine, Shara, \& Rich(2003)]{LSR03a}
L\'epine, S., Shara, M. M., \& Rich, R. M. 2003, \apjl, {\it in press}

\bibitem[L\'epine, Rich, \& Shara(2003)]{LRS03b}
L\'epine, S., Rich, R. M., \& Shara, M. M. 2003, \aj, {\it in press}


\bibitem[Luyten (1979a)]{L79a}
Luyten W. J. 1979, LHS Catalogue: a catalogue of stars
with proper motions exceeding 0.5" annually, University of Minnesota,
Minneapolis ({\it CDS-ViZier catalog number I/87B})

\bibitem[Luyten (1979b)]{L79b}
Luyten W. J. 1979, New Luyten Catalogue of stars with
proper motions larger than two tenths of an arcsecond (NLTT),
University of Minnesota, Minneapolis ({\it CDS-ViZier catalog number
I/98A})

\bibitem[Monet {\it et al.}(2000)]{MFLCHR00}
Monet, D. G., Fisher, M. D., Liebert, J., Canzian, B., Harris, H. C.,
\& Reid, I. N. 2000, \aj, 120, 1541

\bibitem[Monet {\it et al.}(2003)]{Metal03}
Monet, D. G., {\it et al.} 2003, \aj, 125, 984

\bibitem[Perryman(1997)]{P97}
Perryman, M. A. C. 1997, The Hipparcos and Tycho
catalogues. Astrometric and photometric star catalogues derived from
the ESA Hipparcos Space Astrometry Mission, Publisher: Noordwijk,
Netherlands: ESA Publicat

\bibitem[Pokorny, Jones, \& Hambly(2003)]{PJH03}
Pokorny, R. S., Jones, H. R. A., \& Hambly, N. C. 2003, \aap, 397, 584

\bibitem[Reid {\it et al.}(2002)]{RKLGCM02}
Reid, I. Neill, Kirkpatrick, J. Davy, Liebert, James, Gizis, J. E.,
Dahn, C. C., \& Monet, D. G. 2002, \aj, 124, 519

\bibitem[Ruiz {\it et al.}(2001)]{RWRG01}
Ruiz, M. T., Wischnjewsky, M., Rojo, P. M., \& Gonzalez, L. E. 2001,
\apjs, 133, 119

\bibitem[Salim \& Gould(2002)]{SG02}
Salim, S., \& Gould, A. 2002, \apjl, 575, L83

\bibitem[Sanduleak \& Pesch(1988)]{SP88}
Sanduleak, N., \& Pesch, P. 1988, \apjs, 66, 387

\bibitem[Scholz {\it et al.}(2000)]{SIIJM00}
Scholz, R.-D., Irwin, M., Ibata, R., Jahreiss, H., \& Malkov,
O. Yu. 2000, \aap, 353, 958

\end{thebibliography}
\end{document}